\documentclass[5p,twocolumn]{elsarticle}

\usepackage{amsmath,amssymb}
\newcommand{\eqdef}{\stackrel{\text{def}}{=}}
\newcommand{\n}{\nonumber \\}
\newcommand{\bm}{\boldsymbol}
\newcommand{\ignore}[1]{}

\allowdisplaybreaks[4]

\journal{Physics Letters B}

\begin{document}

\begin{frontmatter}

\title{Infinitely many shape invariant discrete quantum mechanical
systems and\\ new exceptional orthogonal polynomials related to\\ the
Wilson and Askey-Wilson polynomials}

\author[SO]{Satoru Odake\corref{cSO}}
\ead{odake@azusa.shinshu-u.ac.jp}
\author[RS]{Ryu Sasaki}
\address[SO]{Department of Physics, Shinshu University,
     Matsumoto 390-8621, Japan}
\address[RS]{Yukawa Institute for Theoretical Physics,
     Kyoto University, Kyoto 606-8502, Japan}
\cortext[cSO]{Corresponding author.}

\begin{abstract}
Two sets of infinitely many exceptional orthogonal polynomials related to
the Wilson and Askey-Wilson polynomials are presented. They are derived
as the eigenfunctions of shape invariant and thus exactly solvable
quantum mechanical Hamiltonians, which are deformations of those for
the Wilson and Askey-Wilson polynomials in terms of a degree $\ell$
($\ell=1,2,\ldots$) eigenpolynomial. These polynomials are exceptional
in the sense that they start from degree $\ell\ge1$ and thus
not constrained by any generalisation of Bochner's theorem.
\end{abstract}




\end{frontmatter}

\section{Introduction}
\label{sec:intro}
\setcounter{equation}{0}

Orthogonal polynomials have played important roles \cite{szego,askey}
on many different stages of mathematics and physics.
For example, the matrix models are one of the recent active stages
for them in particle physics.
As is well-known there are infinitely many orthogonal polynomials,
since they can be constructed by Gram-Schmidt orthonormalisation of
the basis $\{1, x, x^2, \ldots\}$ with respect to any given weight
(or distribution) $dw(x)$, having finite moments
$\int x^ndw(x)<\infty$, $n\in\mathbb{Z}_+$.
A more practical way of generating orthogonal polynomials $\{p_n(x)\}$
is through the three term recurrence relation (Favard's theorem
\cite{Chihara})
\begin{equation}
  xp_n(x)=A_np_{n+1}(x)+B_np_n(x)+C_np_{n-1}(x)\ \ (n\geq 0),
  \label{threeterm}
\end{equation}
with $p_{-1}(x)=0$. Here the coefficients $A_n$, $B_n$ and $C_n$
are real and $A_{n-1}C_n>0$ ($n\geq 1$).
Conversely all the orthogonal polynomials starting with degree 0
satisfy the above three term recurrence relation.
Among many orthogonal polynomials, a central role is played by those
satisfying (second order) differential or difference equations,
in particular those appearing in quantum mechanics or in other
dynamical problems.
According to Bochner \cite{bochner}, all polynomial solutions of
the second order differential equation
\begin{equation}
  \sigma(x)y''+\tau(x)y'=\lambda_ny,
\end{equation}
are {\em classical orthogonal polynomials\/}, that is the Hermite,
Laguerre, Jacobi and Bessel polynomials. Here, $\sigma(x)$ and
$\tau(x)$ are polynomials in $x$ of deg$(\sigma)\le2$, deg$(\tau)\le1$.
On top of these classical orthogonal polynomials, a variety of orthogonal 
polynomials, the so-called ($q$-)Askey scheme of hypergeometric
orthogonal polynomials \cite{askey, nikiforov, ismail, koeswart},
are known to satisfy second order {\em difference\/} equations with
pure imaginary or real shifts. They also satisfy restrictions due to
various generalisation of Bochner's theorem \cite{askeywil}.

In 2008, Gomez-Ullate et al \cite{gomez} introduced two new
orthogonal polynomials satisfying second order differential equations,
which are called exceptional ($X_1$) Laguerre and Jacobi polynomials.
They are consistent with Bochner's theorem, since the polynomials
start with degree 1 rather than degree 0 constant term.
Later reformulation within the framework of quantum mechanics and
shape invariance followed in \cite{quesne, BQR}.
Recently two sets of infinitely many orthogonal polynomials satisfying
second order differential equations are introduced by the present
authors \cite{os16}.
These new polynomials could be called $X_\ell$ Laguerre and Jacobi
polynomials for each positive integer $\ell$. The $X_\ell$ polynomials
start with degree $\ell$ and lack the members from degree 0 to degree
$\ell-1$ but they form a complete basis. The $\ell=1$ examples are
those reported earlier \cite{gomez, quesne, BQR}.

In this letter we present two new sets of infinitely many orthogonal
polynomials satisfying second order {\em difference\/} equations.
They could be called exceptional ($X_\ell$) Wilson and Askey-Wilson
polynomials, since they start with degree $\ell\ge1$.
They are complete and satisfy various restrictions due to various
generalisation of Bochner's theorem \cite{askeywil}.
It is quite natural to expect that these exceptional polynomials
(Laguerre, Jacobi, Wilson and Askey-Wilson) will find fruitful
applications in many branches of physics and mathematics.
The method of deriving these new polynomials is, as in the cases of
the $X_\ell$ Laguerre and Jacobi polynomials, {\em shape invariance\/}
\cite{genden} in quantum mechanics.
These polynomials are obtained as the main part of the eigenfunctions
of a corresponding Hamiltonian of `discrete' quantum mechanics (dQM).
Thus the orthogonality and completeness of the polynomials are a priori
guaranteed. Shape invariance combined with Crum's method \cite{crum}
or the factorisation method \cite{infhul} or the supersymmetric
quantum mechanics \cite{susyqm} is a sufficient condition for exact
solvability of a quantum mechanical Hamiltonian;
the entire energy spectrum and the corresponding eigenfunctions can be
obtained exactly in an algebraic way.
Discrete quantum mechanics (dQM) was introduced in 2004 by the present
authors \cite{os4,os5} in order to reformulate the theory of
hypergeometric orthogonal polynomials \cite{askey, ismail,koeswart}
within the framework of quantum mechanics, so that various concepts
and methods of QM would be available for the theory of these orthogonal
polynomials. This revealed the shape invariance of the Hamiltonians
for all the hypergeometric orthogonal polynomials \cite{koeswart},
ranging from the Meixner-Pollaczek, continuous (dual) Hahn, Wilson,
Askey-Wilson and ($q$-)Racah together with their reductions
\cite{os12,os13}. Moreover all these dQM systems turn out to be
exactly solvable in the Heisenberg picture, too \cite{os7-8}.
The positive/negative energy parts of the Heisenberg operator solution
define the annihilation/creation operators, which form dynamical
symmetry algebras together with the Hamiltonian; for example,
the $q$-oscillator algebra \cite{os11}.
Another benefit of dQM is that it provided many examples of exactly
solvable birth and death processes \cite{bdsol}.
The Hamiltonians for the two infinite sets of exceptional orthogonal
polynomials are obtained by deforming those for the Wilson and
Askey-Wilson polynomials in terms of a degree $\ell$ eigenpolynomial
with {\em twisted\/} arguments, so that no singularity would appear
in the domains where the orthogonality measures are defined.
This is essentially the same procedure employed for the exceptional
Laguerre and Jacobi polynomials \cite{os16}.
The necessary technology, the discrete versions of Crum's theorem,
was published recently \cite{os15}.

Here we present our preliminary results on the two sets of infinitely
many exceptional orthogonal polynomials and their discrete quantum
mechanical formulation, without proof.
In the next section, the general setting of dQM and the method of
shape invariance are recapitulated.
In section three the relationship between the original shape invariance
and its multiplicative deformation for the generation of exceptional
polynomials is explained for two typical cases of the Wilson and
Askey-Wilson polynomials.
Sections four and five provide the detailed data of the exceptional
Wilson and Askey-Wilson polynomials, respectively.
The final section is for a summary and comments.

\section{General setting: shape invariance}
\label{sec:settings}

In this letter we consider one dimensional discrete quantum mechanical
systems with pure imaginary shifts.
The dynamical variables are the real coordinate $x$ ($x_1<x<x_2$) and
the conjugate momentum $p=-i\partial_x$, which are governed by the
following generic Hamiltonian:
\begin{align}
  &\mathcal{H}\eqdef\sqrt{V(x)}\,e^{\gamma p}\sqrt{V^*(x)}
  +\!\sqrt{V^*(x)}\,e^{-\gamma p}\sqrt{V(x)}\n
  &\phantom{\mathcal{H}\eqdef\ \ }
  -V(x)-V^*(x)=\mathcal{A}^{\dagger}\mathcal{A},
  \label{discrham}\\
  &\mathcal{A}\eqdef i\bigl(e^{\frac{\gamma}{2}p}\sqrt{V^*(x)}
  -e^{-\frac{\gamma}{2}p}\sqrt{V(x)}\,\bigr),\n
  &\mathcal{A}^{\dagger}\eqdef -i\bigl(\sqrt{V(x)}\,e^{\frac{\gamma}{2}p}
  -\sqrt{V^*(x)}\,e^{-\frac{\gamma}{2}p}\bigr).
\end{align}
It is factorised and positive semi-definite.
Here the potential function $V(x)$ is an analytic function of $x$ and
$\gamma$ is a real constant to be specified for each specific case,
see for example \eqref{etaWilson}-\eqref{etaAWilson} for the $\gamma$
and the domain $x_1<x<x_2$ for the two cases discussed in this letter.
The $*$-operation on an analytic function $f(x)=\sum_na_nx^n$
($a_n\in\mathbb{C}$) is defined as $f^*(x)=\sum_na_n^*x^n$, in which
$a_n^*$ is the complex conjugation of $a_n$.
Since the momentum operator appears in exponentiated forms, the
Schr\"{o}dinger equation
$\mathcal{H}\phi_n(x)=\mathcal{E}_n\phi_n(x)$
\ $(n=0,1,2,\ldots)$,
is an analytic difference equation with pure imaginary shifts instead
of a differential equation.
The groundstate is annihilated by $\mathcal{A}$, $\mathcal{A}\phi_0=0$
($\Rightarrow$ $\mathcal{H}\phi_0=0$) and can be chosen `real',
$\phi_0^*(x)=\phi_0(x)$.
Throughout this letter we consider those systems which have a
square-integrable groundstate together with an infinite number of
discrete energy levels:
$0=\mathcal{E}_0 <\mathcal{E}_1 < \mathcal{E}_2 < \cdots$.

{\em Shape invariance\/}, a sufficient condition for exact solvability
\cite{genden}, is realised by specific dependence of the potential
function on a set of parameters
$\bm{\lambda}=(\lambda_1,\lambda_2,\ldots)$, to be denoted by
$V(x;\bm{\lambda})$, $\mathcal{A}(\bm{\lambda})$,
$\mathcal{H}(\bm{\lambda})$, $\mathcal{E}_n(\bm{\lambda})$,
$\phi_n(x;\bm{\lambda})$ etc.
The shape invariance condition to be discussed in this letter is
\cite{os4,os5,os6,os13}
\begin{equation}
  \mathcal{A}(\bm{\lambda})\mathcal{A}(\bm{\lambda})^{\dagger}
  =\kappa\mathcal{A}(\bm{\lambda}+\bm{\delta})^{\dagger}
  \mathcal{A}(\bm{\lambda}+\bm{\delta})
  +\mathcal{E}_1(\bm{\lambda}),
\end{equation}
in which $\bm{\delta}$ is a certain shift of the parameters and
$\kappa$ is a positive constant. This condition is equivalent to
the following two relations:
\begin{align}
  &V(x-i\tfrac{\gamma}{2};\bm{\lambda})
  V^*(x-i\tfrac{\gamma}{2};\bm{\lambda})\n
  &\qquad
  =\kappa^2\,V(x;\bm{\lambda}+\bm{\delta})
  V^*(x-i\gamma;\bm{\lambda}+\bm{\delta}),
  \label{shapeinv2a}\\
  &V(x+i\tfrac{\gamma}{2};\bm{\lambda})
  +V^*(x-i\tfrac{\gamma}{2};\bm{\lambda})\n
  &\qquad
  =\kappa\bigl(V(x;\bm{\lambda}+\bm{\delta})
  +V^*(x;\bm{\lambda}+\bm{\delta})\bigr)
  -\mathcal{E}_1(\bm{\lambda}).
  \label{shapeinv2b}
\end{align}
Then the entire set of discrete eigenvalues and the corresponding
eigenfunctions of $\mathcal{H}=\mathcal{H}(\bm{\lambda})$,
\begin{equation}
  \mathcal{H}(\bm{\lambda})\phi_n(x;\bm{\lambda})
  =\mathcal{E}_n(\bm{\lambda})\phi_n(x;\bm{\lambda}),
\end{equation}
is determined algebraically:
\begin{align}
  &\mathcal{E}_n(\bm{\lambda})=\sum_{s=0}^{n-1}
  \kappa^s\mathcal{E}_1(\bm{\lambda}+s\bm{\delta}),\\
  &\phi_n(x;\bm{\lambda})\propto
  \mathcal{A}(\bm{\lambda})^{\dagger}
  \mathcal{A}(\bm{\lambda}+\bm{\delta})^{\dagger}
  \cdots
  \mathcal{A}(\bm{\lambda}+(n-1)\bm{\delta})^{\dagger}\n
  &\qquad\qquad\qquad\qquad\qquad\qquad\qquad\times
  \phi_0(x;\bm{\lambda}+n\bm{\delta}).
  \label{phin=AA..phi0}
\end{align}

\section{Deformation of shape invariant systems}
\label{sec:shape}

In \S\ref{sec:shape_org} we recapitulate two known shape invariant
systems, whose eigenfunctions are described by the Wilson polynomial
and the Askey-Wilson polynomial.
In \S\ref{sec:shape_new} we deform them in terms of a degree $\ell$
eigenpolynomial ($\ell=1,2,\ldots$) and  present a new shape invariant
system. The general structure is common for these two cases.
Explicit expressions of  various quantities will be given in
\S\ref{sec:W} and \S\ref{sec:AW}, respectively.

\subsection{original shape invariant systems}
\label{sec:shape_org}

In \cite{os13} various examples of exactly solvable dQM systems are
explored with emphasis on shape invariance and closure relation.
A closure relation is another sufficient condition for exact solvability,
with which the Heisenberg operator solution for the sinusoidal
coordinate $\eta(x)$ can be obtained exactly \cite{os7-8} (see \cite{os14}
for further developments).
The examples in \cite{os13} are arranged by the sinusoidal coordinates
($\eta(x)=x,x^2,\cos x$) and the Wilson and Askey-Wilson polynomials
are the most generic members for $\eta(x)=x^2$ and $\cos x$, respectively,
\begin{alignat}{5}
  \text{Wilson}:&\ &\eta(x)&=x^2,&&0<x<\infty,&\ \ &&\gamma&=1,
  \label{etaWilson}\\
  \!\!\!\text{Askey-Wilson}:&&\eta(x)&=\cos x,\ \ &&0<x<\pi,&&&
  \gamma&=\log q.
  \label{etaAWilson}
\end{alignat}
They contain four parameters
$\bm{\lambda}=(\lambda_1,\lambda_2,\lambda_3,\lambda_4)$
(and one more parameter $q$ ($0<q<1$) in the Askey-Wilson case)
and the data for shape invariance are
\begin{equation}
  \bm{\delta}=(\tfrac12,\tfrac12,\tfrac12,\tfrac12),\quad
  \kappa=\Bigl\{
  \begin{array}{ll}
  1&\!\!\!\text{: Wilson},\\
  q^{-1}&\!\!\!\text{: Askey-Wilson}.
  \end{array}
\end{equation}
It is straightforward to verify that the potential function
$V(x;\bm{\lambda})$, \eqref{VWilson} in \S\ref{sec:W} and
\eqref{VAWilson} in \S\ref{sec:AW}, satisfies the shape invariance
conditions \eqref{shapeinv2a}--\eqref{shapeinv2b}.

The groundstate is annihilated by $\mathcal{A}$
($\mathcal{A}\phi_0=0$), namely,
\begin{align}
  &\sqrt{V^*(x-i\tfrac{\gamma}{2};\bm{\lambda})}\,
  \phi_0(x-i\tfrac{\gamma}{2};\bm{\lambda})\n
  &\qquad\qquad
  =\sqrt{V(x+i\tfrac{\gamma}{2};\bm{\lambda})}\,
  \phi_0(x+i\tfrac{\gamma}{2};\bm{\lambda}),
\end{align}
and it satisfies also
\begin{equation}
  \phi_0(x;\bm{\lambda}+\bm{\delta})
  =\varphi(x)\sqrt{V(x+i\tfrac{\gamma}{2};\bm{\lambda})}\,
  \phi_0(x+i\tfrac{\gamma}{2};\bm{\lambda}),
\end{equation}
where an auxiliary function $\varphi(x)$ is defined by
\begin{equation}
  \varphi(x)\eqdef\Bigl\{
  \begin{array}{ll}
  2x&\!\!\!\text{: Wilson},\\
  2\sin x&\!\!\!\text{: Askey-Wilson}.
  \end{array}
\end{equation}
The eigenfunctions $\{\phi_n\}$ have the following form:
\begin{equation}
  \phi_n(x;\bm{\lambda})
  =\phi_0(x;\bm{\lambda})P_n\bigl(\eta(x);\bm{\lambda}\bigr),
\end{equation}
where $P_n(x;\bm{\lambda})$ is a polynomial of degree $n$
in $x$ and is `real', $P_n^*(x)=P_n(x)$.
In the two cases we are considering,  $P_n$ is the Wilson polynomial
or the Askey-Wilson polynomial, which are orthogonal with respect to
the weight function $\phi_0(x;\bm{\lambda})^2$:
\begin{equation}
  \int_{x_1}^{x_2}\phi_0(x;\bm{\lambda})^2
  P_n\bigl(\eta(x);\bm{\lambda}\bigr)P_m\bigl(\eta(x);\bm{\lambda}\bigr)dx
  =h_n(\bm{\lambda})\delta_{nm}.
\end{equation}
See \S\ref{sec:W} and \S\ref{sec:AW} for the explicit forms of
the groundstate wavefunction $\phi_0(x;\bm{\lambda})$, the
definition of the polynomial $P_n\bigl(\eta(x);\bm{\lambda}\bigr)$
and the normalisation constant $h_n(\bm{\lambda})$.

The action of the operators $\mathcal{A}$ and $\mathcal{A}^{\dagger}$
on the eigenfunction $\phi_n$ is
\begin{align}
  &\mathcal{A}(\bm{\lambda})
  \phi_n(x;\bm{\lambda})
  =f_n(\bm{\lambda})\phi_{n-1}(x;\bm{\lambda}+\bm{\delta}),\\
  &\mathcal{A}(\bm{\lambda})^{\dagger}
  \phi_{n-1}(x;\bm{\lambda}+\bm{\delta})
  =b_{n-1}(\bm{\lambda})\phi_n(x;\bm{\lambda}),
\end{align}
in which $f_n(\bm{\lambda})$ and $b_{n-1}(\bm{\lambda})$ are real
constants related to $\mathcal{E}_n$ as
$f_n(\bm{\lambda})b_{n-1}(\bm{\lambda})=\mathcal{E}_n(\bm{\lambda})$.
Their explicit forms are given in \eqref{fbWilson} in \S\ref{sec:W} and
in \eqref{fbAWilson} in \S\ref{sec:AW}.
The forward shift operator $\mathcal{F}$ and the backward shift operator
$\mathcal{B}$ are defined by removing the effects of the groundstate
wavefunctions
\begin{align}
  &\mathcal{F}(\bm{\lambda})\eqdef
  \phi_0(x;\bm{\lambda}+\bm{\delta})^{-1}\circ
  \mathcal{A}(\bm{\lambda})\circ\phi_0(x;\bm{\lambda})\n
  &\phantom{\mathcal{F}(\bm{\lambda})}=
  i\varphi(x)^{-1}\bigl(e^{\frac{\gamma}{2}p}-e^{-\frac{\gamma}{2}p}\bigr),
  \label{defforward}\\
  &\mathcal{B}(\bm{\lambda})\eqdef
  \phi_0(x;\bm{\lambda})^{-1}\circ
  \mathcal{A}(\bm{\lambda})^{\dagger}\circ
  \phi_0(x;\bm{\lambda}+\bm{\delta})\n
  &\phantom{\mathcal{B}(\bm{\lambda})}=
  -i\bigl(V(x;\bm{\lambda})e^{\frac{\gamma}{2}p}
  -V^*(x;\bm{\lambda})e^{-\frac{\gamma}{2}p}\bigr)\varphi(x),
  \label{defbackward}
\end{align}
and their action on the polynomial $P_n(\eta;\bm{\lambda})$ is
\begin{align}
  &\mathcal{F}(\bm{\lambda})
  P_n\bigl(\eta(x);\bm{\lambda}\bigr)
  =f_n(\bm{\lambda})P_{n-1}\bigl(\eta(x);\bm{\lambda}+\bm{\delta}\bigr),
  \label{forwardaction}\\
  &\mathcal{B}(\bm{\lambda})
  P_{n-1}\bigl(\eta(x);\bm{\lambda}+\bm{\delta}\bigr)
  =b_{n-1}(\bm{\lambda})P_n\bigl(\eta(x);\bm{\lambda}\bigr).
  \label{backwardaction}
\end{align}
The difference equation for the polynomials reads
\begin{align}
  &\widetilde{\mathcal{H}}(\bm{\lambda})\eqdef
  \mathcal{B}(\bm{\lambda})\mathcal{F}(\bm{\lambda}),\\
  &\widetilde{\mathcal{H}}(\bm{\lambda})P_n\bigl(\eta(x);\bm{\lambda}\bigr)
  =\mathcal{E}_n(\bm{\lambda})P_n\bigl(\eta(x);\bm{\lambda}\bigr),\\
  \text{or:}&\quad\ V(x)\bigl(P_n\bigl(\eta(x-i\gamma)\bigr)
  -P_n\bigl(\eta(x)\bigr)\bigr)\n
  &\quad
  +V^*(x)\bigl(P_n\bigl(\eta(x+i\gamma)\bigr)-P_n\bigl(\eta(x)\bigr)\bigr)\n
  &\quad =\mathcal{E}_nP_n\bigl(\eta(x)\bigr).
  \label{difeq}
\end{align}
In the last equation \eqref{difeq}, the explicit parameter $\bm{\lambda}$
dependence was suppressed for simplicity.
The above formulas for the polynomials \eqref{defforward}--\eqref{difeq}
are well known \cite{koeswart}.

\subsection{new shape invariant systems}
\label{sec:shape_new}

By deforming the shape invariant systems in the previous subsection
in terms of a degree $\ell$ eigenpolynomial, we present an infinite
number of shape invariant systems indexed by a non-negative integer
$\ell=0,1,2,\ldots$. For $\ell=0$, it is the original one.
Explicit expressions of various quantities ($a_{\ell,n,1}$,
$a_{\ell,n,2}$, $b_{\ell,n,1}$, $h_{\ell,n}$) will be given in
\S\ref{sec:W} and \S\ref{sec:AW}.

In terms of a twist operation $\mathfrak{t}$,
\begin{equation}
  \bm{\lambda}=(\lambda_1,\lambda_2,\lambda_3,\lambda_4),\quad
  \mathfrak{t}(\bm{\lambda})\eqdef
  (-\lambda_1,-\lambda_2,\lambda_3,\lambda_4),
\end{equation}
we define a degree $\ell$ polynomial $\xi_{\ell}(x)$ from the
eigenpolynomial $P_\ell(x)$:
\begin{equation}
  \xi_{\ell}(x;\bm{\lambda})\eqdef P_{\ell}\bigl(x;
  \mathfrak{t}\bigl(\bm{\lambda}+(\ell-1)\bm{\delta}\bigr)\bigr).
\end{equation}
Then we define a potential function $V_{\ell}(x;\bm{\lambda})$
and a Hamiltonian $\mathcal{H}_{\ell}(\bm{\lambda})$:
\begin{align}
  &V_{\ell}(x;\bm{\lambda})\eqdef V(x;\bm{\lambda}+\ell\bm{\delta})\,
  \frac{\xi_{\ell}(\eta(x+i\frac{\gamma}{2});\bm{\lambda})}
  {\xi_{\ell}(\eta(x-i\frac{\gamma}{2});\bm{\lambda})}\n
  &\qquad\qquad\qquad\qquad\qquad\times
  \frac{\xi_{\ell}(\eta(x-i\gamma);\bm{\lambda}+\bm{\delta})}
  {\xi_{\ell}(\eta(x);\bm{\lambda}+\bm{\delta})},\\
  &V_{\ell}^*(x;\bm{\lambda})=V^*(x;\bm{\lambda}+\ell\bm{\delta})\,
  \frac{\xi_{\ell}(\eta(x-i\frac{\gamma}{2});\bm{\lambda})}
  {\xi_{\ell}(\eta(x+i\frac{\gamma}{2});\bm{\lambda})}\n
  &\qquad\qquad\qquad\qquad\qquad\times
  \frac{\xi_{\ell}(\eta(x+i\gamma);\bm{\lambda}+\bm{\delta})}
  {\xi_{\ell}(\eta(x);\bm{\lambda}+\bm{\delta})},\\
  &\mathcal{A}_{\ell}(\bm{\lambda})\eqdef
  i\bigl(e^{\frac{\gamma}{2}p}\sqrt{V_{\ell}^*(x;\bm{\lambda})}
  -e^{-\frac{\gamma}{2}p}\sqrt{V_{\ell}(x;\bm{\lambda})}\,\bigr),\\
  &\mathcal{A}_{\ell}(\bm{\lambda})^{\dagger}\eqdef
  -i\bigl(\sqrt{V_{\ell}(x;\bm{\lambda})}\,e^{\frac{\gamma}{2}p}
  -\sqrt{V_{\ell}^*(x;\bm{\lambda})}\,e^{-\frac{\gamma}{2}p}\bigr),\\
  &\mathcal{H}_{\ell}(\bm{\lambda})\eqdef
  \mathcal{A}_{\ell}(\bm{\lambda})^{\dagger}
  \mathcal{A}_{\ell}(\bm{\lambda}).
\end{align}
It is interesting to note that the deformation of the potential function
$V(x)$ is multiplicative in contrast to the additive deformation
of the prepotential $w(x)$ in the ordinary QM \cite{os16}.
With an appropriate choice of the range of parameters, the polynomials
$\xi_{\ell}\bigl(\eta(x);\bm{\lambda}\bigr)$ and
$\xi_{\ell}\bigl(\eta(x);\bm{\lambda}+\bm{\delta}\bigr)$ have no zero in
the rectangular domain in the complex $x$ plane,
$x_1\le \text{Re}\,x\le x_2$,
$-\frac{|\gamma|}{2}\le \text{Im}\,x\le \frac{|\gamma|}{2}$
and the Hamiltonian $\mathcal{H}_{\ell}$ is well-defined and hermitian
(self-adjoint).
The potential function $V_{\ell}(x)$ satisfies the shape invariance
conditions \eqref{shapeinv2a}--\eqref{shapeinv2b},
\begin{align}
  &V_{\ell}(x-i\tfrac{\gamma}{2};\bm{\lambda})
  V_{\ell}^*(x-i\tfrac{\gamma}{2};\bm{\lambda})\n
  &\qquad
  =\kappa^2 V_{\ell}(x;\bm{\lambda}+\bm{\delta})
  V_{\ell}^*(x-i\gamma;\bm{\lambda}+\bm{\delta}),\\
  &V_{\ell}(x+i\tfrac{\gamma}{2};\bm{\lambda})
  +V_{\ell}^*(x-i\tfrac{\gamma}{2};\bm{\lambda})\n
  &\qquad
  =\kappa\bigl(V_{\ell}(x;\bm{\lambda}+\bm{\delta})
  +V_{\ell}^*(x;\bm{\lambda}+\bm{\delta})\bigr)
  -\mathcal{E}_{\ell,1}(\bm{\lambda}).
\end{align}
By using \eqref{phin=AA..phi0} as a Rodrigues type formula,
we obtain the complete set of eigenfunctions ($n=0,1,2,\ldots$):
\begin{align}
  &\mathcal{H}_{\ell}(\bm{\lambda})\phi_{\ell,n}(x;\bm{\lambda})
  =\mathcal{E}_{\ell,n}(\bm{\lambda})\phi_{\ell,n}(x;\bm{\lambda}),\\
  &\mathcal{E}_{\ell,n}(\bm{\lambda})
  =\mathcal{E}_n(\bm{\lambda}+\ell\bm{\delta}),\\
  &\phi_{\ell,0}(x;\bm{\lambda})=
  \frac{\phi_0(x;\bm{\lambda}+\ell\bm{\delta})}
  {|\xi_{\ell}(\eta(x-i\frac{\gamma}{2});\bm{\lambda})|}
  \,\xi_{\ell}\bigl(\eta(x);\bm{\lambda}+\bm{\delta}\bigr),\\
  &\psi_{\ell}(x;\bm{\lambda})\eqdef
  \frac{\phi_0(x;\bm{\lambda}+\ell\bm{\delta})}
  {|\xi_{\ell}(\eta(x-i\frac{\gamma}{2});\bm{\lambda})|},\\
  &\phi_{\ell,n}(x;\bm{\lambda})=\psi_{\ell}(x;\bm{\lambda})
  P_{\ell,n}\bigl(\eta(x);\bm{\lambda}\bigr).
\end{align}
Here a degree $\ell+n$ polynomial in $x$, the polynomial eigenfunction
$P_{\ell,n}(x;\bm{\lambda})$
is given by
\begin{align}
  &P_{\ell,n}(x;\bm{\lambda})\eqdef
  a_{\ell,n}(x;\bm{\lambda})P_n(x;\bm{\lambda}+\ell\bm{\delta})\n
  &\qquad\qquad\qquad
  +b_{\ell,n}(x;\bm{\lambda})P_{n-1}(x;\bm{\lambda}+\ell\bm{\delta}),\\
  &a_{\ell,n}(x;\bm{\lambda})\eqdef
  \xi_{\ell}(x;\bm{\lambda}+\bm{\delta})\n
  &\qquad\qquad\qquad
  +a_{\ell,n,1}(\bm{\lambda})
  \xi_{\ell-1}(x;\bm{\lambda}+\bm{\delta}+\bm{\delta'})\n
  &\qquad\qquad\qquad
  +a_{\ell,n,2}(\bm{\lambda})
  \xi_{\ell-2}(x;\bm{\lambda}+2\bm{\delta}+\bm{\delta'}),\\
  &b_{\ell,n}(x;\bm{\lambda})\eqdef
  b_{\ell,n,1}(\bm{\lambda})
  \xi_{\ell-1}(x;\bm{\lambda}+\bm{\delta}+\bm{\delta'}),
\end{align}
where $\bm{\delta}'$ is
\begin{equation}
  \bm{\delta'}\eqdef\mathfrak{t}(\bm{\delta})
  =(-\tfrac12,-\tfrac12,\tfrac12,\tfrac12).
\end{equation}
Here the coefficients $a_{\ell,n,1}(\bm{\lambda})$,
$a_{\ell,n,2}(\bm{\lambda})$ and $b_{\ell,n,1}(\bm{\lambda})$ are real.
Although the polynomial eigenfunction
$P_{\ell,n}\bigl(\eta(x);\bm{\lambda}\bigr)$ is a degree
$\ell+n$ polynomial in $\eta(x)$, it has only $n$ zeros in the
domain $x_1<x<x_2$. These polynomials are orthogonal with respect
to the weight function $\psi_{\ell}(x;\bm{\lambda})^2$:
\begin{equation}
  \int_{x_1}^{x_2}\!\!\psi_{\ell}(x;\bm{\lambda})^2
  P_{\ell,n}\bigl(\eta(x);\bm{\lambda}\bigr)
  P_{\ell,m}\bigl(\eta(x);\bm{\lambda}\bigr)dx
  =h_{\ell,n}(\bm{\lambda})\delta_{nm}.
  \label{hln}
\end{equation}
They form a complete basis of the Hilbert space just like the Wilson
or the Askey-Wilson polynomials $P_n\bigl(\eta(x);\bm{\lambda}\bigr)$
in the $\ell=0$ case.
It should be stressed that $\psi_{\ell}(x;\bm{\lambda})$ is not
annihilated by $\mathcal{A}_{\ell}$.

The action of the operators $\mathcal{A}_{\ell}$ and
$\mathcal{A}_{\ell}^{\dagger}$ on the eigenfunction $\phi_{\ell,n}$ is
\begin{align}
  &\mathcal{A}_{\ell}(\bm{\lambda})\phi_{\ell,n}(x;\bm{\lambda})
  =f_{\ell,n}(\bm{\lambda})\phi_{\ell,n-1}(x;\bm{\lambda}+\bm{\delta}),\\
  &\mathcal{A}_{\ell}(\bm{\lambda})^{\dagger}
  \phi_{\ell,n-1}(x;\bm{\lambda}+\bm{\delta})
  =b_{\ell,n-1}(\bm{\lambda})\phi_{\ell,n}(x;\bm{\lambda}),\\
  &f_{\ell,n}(\bm{\lambda})=f_n(\bm{\lambda}+\ell\bm{\delta}),\quad
  b_{\ell,n}(\bm{\lambda})=b_n(\bm{\lambda}+\ell\bm{\delta}).
\end{align}
The forward shift operator $\mathcal{F}_{\ell}$ and the backward shift
operator $\mathcal{B}_{\ell}$ are defined in a similar way as before
\begin{align}
  &\mathcal{F}_{\ell}(\bm{\lambda})\eqdef
  \psi_{\ell}(x;\bm{\lambda}+\bm{\delta})^{-1}\circ
  \mathcal{A}_{\ell}(\bm{\lambda})\circ
  \psi_{\ell}(x;\bm{\lambda})\n
  &\quad=
  \frac{i}{\varphi(x)\xi_{\ell}(\eta(x);\bm{\lambda})}
  \Bigl(\xi_{\ell}\bigl(\eta(x+i\tfrac{\gamma}{2});\bm{\lambda}
  +\bm{\delta}\bigr)e^{\frac{\gamma}{2}p}\n
  &\qquad\qquad\qquad\qquad
  -\xi_{\ell}\bigl(\eta(x-i\tfrac{\gamma}{2});\bm{\lambda}
  +\bm{\delta}\bigr)e^{-\frac{\gamma}{2}p}\Bigr),\\
  &\mathcal{B}_{\ell}(\bm{\lambda})\eqdef
  \psi_{\ell}(x;\bm{\lambda})^{-1}\circ
  \mathcal{A}_{\ell}(\bm{\lambda})^{\dagger}\circ
  \psi_{\ell}(x;\bm{\lambda}+\bm{\delta})\n
  &\quad=
  \frac{-i}{\xi_{\ell}(\eta(x);\bm{\lambda}+\bm{\delta})}
  \Bigl(V(x;\bm{\lambda}+\ell\bm{\delta})
  \xi_{\ell}\bigl(\eta(x+i\tfrac{\gamma}{2});\bm{\lambda}\bigr)
  e^{\frac{\gamma}{2}p}\n
  &\qquad
  -V^*(x;\bm{\lambda}+\ell\bm{\delta})
  \xi_{\ell}\bigl(\eta(x-i\tfrac{\gamma}{2});\bm{\lambda}\bigr)
  e^{-\frac{\gamma}{2}p}\Bigr)\varphi(x),
\end{align}
and their action on the polynomial $P_{\ell,n}(\eta;\bm{\lambda})$ is
\begin{align}
  &\mathcal{F}_{\ell}(\bm{\lambda})
  P_{\ell,n}\bigl(\eta(x);\bm{\lambda}\bigr)
  =f_{\ell,n}(\bm{\lambda})
  P_{\ell,n-1}\bigl(\eta(x);\bm{\lambda}+\bm{\delta}\bigr),\\
  &\mathcal{B}_{\ell}(\bm{\lambda})
  P_{\ell,n-1}\bigl(\eta(x);\bm{\lambda}+\bm{\delta}\bigr)
  =b_{\ell,n-1}(\bm{\lambda})P_{\ell,n}\bigl(\eta(x);\bm{\lambda}\bigr).
\end{align}
The operator $\widetilde{\mathcal{H}}_{\ell}(\bm{\lambda})$ acting on the
polynomial eigenfunctions is defined by
\begin{align}
  &\widetilde{\mathcal{H}}_{\ell}(\bm{\lambda})\eqdef
  \mathcal{B}_{\ell}(\bm{\lambda})\mathcal{F}_{\ell}(\bm{\lambda})\n
  &\phantom{\widetilde{\mathcal{H}}_{\ell}(\bm{\lambda})}
  =V(x;\bm{\lambda}+\ell\bm{\delta})
  \frac{ \xi_{\ell}\bigl(\eta(x+i\tfrac{\gamma}{2});\bm{\lambda}\bigr)}
  { \xi_{\ell}\bigl(\eta(x-i\tfrac{\gamma}{2});\bm{\lambda}\bigr)}\n
  &\phantom{\widetilde{\mathcal{H}}_{\ell}(\bm{\lambda})}
  \qquad\times\biggl(e^{\gamma p}
  -\frac{\xi_{\ell}\bigl(\eta(x-i\gamma);\bm{\lambda}+\bm{\delta}\bigr)}
  {\xi_{\ell}\bigl(\eta(x);\bm{\lambda}+\bm{\delta}\bigr)}\biggr)\n
  &\phantom{\widetilde{\mathcal{H}}_{\ell}(\bm{\lambda})}
  \phantom{=}+V^*(x;\bm{\lambda}+\ell\bm{\delta})
  \frac{ \xi_{\ell}\bigl(\eta(x-i\tfrac{\gamma}{2});\bm{\lambda}\bigr)}
  {\xi_{\ell}\bigl(\eta(x+i\tfrac{\gamma}{2});\bm{\lambda}\bigr)}\n
  &\phantom{\widetilde{\mathcal{H}}_{\ell}(\bm{\lambda})}
  \qquad\times \biggl(e^{-\gamma p}
  -\frac{\xi_{\ell}\bigl(\eta(x+i\gamma);\bm{\lambda}+\bm{\delta}\bigr)}
  {\xi_{\ell}\bigl(\eta(x);\bm{\lambda}+\bm{\delta}\bigr)}\biggr),\\
  &\widetilde{\mathcal{H}}_{\ell}(\bm{\lambda})
  P_{\ell,n}\bigl(\eta(x);\bm{\lambda}\bigr)
  =\mathcal{E}_{\ell,n}(\bm{\lambda})
  P_{\ell,n}\bigl(\eta(x);\bm{\lambda}\bigr).
\end{align}

\section{Exceptional Wilson polynomials}
\label{sec:W}

We present the data in \S\ref{sec:shape} for the Wilson polynomial.

\subsection{original shape invariant system}
\label{sec:W_org}

We take the four parameters as follows:
\begin{align}
  &\bm{\lambda}\eqdef(a_1,a_2,a_3,a_4),\quad
  \text{Re}\,a_i>0\ \ (1\leq i\leq 4),\n
  &\{a_1^*,a_2^*,a_3^*,a_4^*\}=\{a_1,a_2,a_3,a_4\} \quad (\text{as a set}).
  \label{para_cond_W}
\end{align}
The potential function, ground state, eigenpolynomial, etc. are
given by
\begin{align}
  &V(x;\bm{\lambda})\eqdef\frac{\prod_{j=1}^4(a_j+ix)}{2ix(2ix+1)},
  \label{VWilson}\\
  &\phi_0(x;\bm{\lambda})\eqdef
  \biggl|\frac{\prod_{j=1}^4\Gamma(a_j+ix)}{\Gamma(2ix)}\biggr|,\\
  &P_n\bigl(\eta(x);\bm{\lambda}\bigr)=W_n(x^2;a_1,a_2,a_3,a_4)
  \eqdef\prod_{j=2}^4(a_1+a_j)_n\n
  &\qquad\quad\times
  {}_4F_3\Bigl(
  \genfrac{}{}{0pt}{}{-n,\,n+b-1,\,a_1+ix,\,a_1-ix}
  {a_1+a_2,\,a_1+a_3,\,a_1+a_4}\Bigm|1\Bigr),\\
  &\mathcal{E}_n(\bm{\lambda})=n\bigl(n+b-1\bigr),
  \quad b\eqdef\sum_{j=1}^4a_j,\\
  &f_n(\bm{\lambda})=-n(n+b-1),
  \quad b_n(\bm{\lambda})=-1,
  \label{fbWilson}\\
  &h_n(\bm{\lambda})=\frac{2\pi n!\,(n+b-1)_n}{\Gamma(2n+b)}
  \!\!\prod_{1\leq j<k\leq 4}\!\!\!\!\Gamma(n+a_j+a_k).
  \label{defhnWilson}
\end{align}
Here $W_n(x^2;a_1,a_2,a_3,a_4)$ is the Wilson polynomial,
which is symmetric with respect to $a_1,\ldots,a_4$.

\subsection{new shape invariant systems}
\label{sec:W_new}

We generically restrict the original parameter range \eqref{para_cond_W}
as follows:
\begin{align}
  &a_1,a_2\in\mathbb{R}, \quad 
  \{a_3^*,a_4^*\}=\{a_3,a_4\} \quad (\text{as a set}),\n
  &0<a_j<\text{Re}\,a_k\ \ (j=1,2;k=3,4).
  \label{range:W}
\end{align}
The polynomial $\xi_{\ell}(x)$ is
\begin{equation}
  \xi_{\ell}(x;\bm{\lambda})=W_{\ell}(x;
  -a_1-\tfrac{\ell-1}{2},-a_2-\tfrac{\ell-1}{2},
  a_3+\tfrac{\ell-1}{2},a_4+\tfrac{\ell-1}{2}),
\end{equation}
which is symmetric under $a_1\leftrightarrow a_2$ and/or
$a_3\leftrightarrow a_4$.
It is elementary to show that $\xi_{\ell}(x;\bm{\lambda})$ and
$\xi_{\ell}(x;\bm{\lambda}+\bm{\delta})$ have no zero on the half
real line $0\le x<\infty$.
The hermiticity of $\mathcal{H}_\ell$ requires a stronger condition
of no zero in the rectangular domain
$-\frac{1}{2}\le \text{Im}\, x\le \frac{1}{2}$.
It can be determined purely algebraically for each $\ell$.
For the lowest case, $\ell=1$, hermiticity is satisfied by
\begin{equation}
  (a_3+a_4)(a_1a_2+\tfrac14)<(a_1+a_2)(a_3a_4+\tfrac14).
  \label{wonerange}
\end{equation}
The higher $\ell$ goes the restriction due to hermiticity becomes
less stringent.

The real coefficients $a_{\ell,n,1}$, $a_{\ell,n,2}$ and $b_{\ell,n,1}$
are
\begin{align}
  &a_{\ell,n,1}(\bm{\lambda})=
  \frac{\ell n(a_1+a_2-a_3-a_4-\ell+1)}
  {a_1+a_2-a_3-a_4-2(\ell-1)}\n
  &\quad\times
  \bigl(a_1+a_2+a_3+a_4+2(n+\ell-1)\bigr)^{-1}\n
  &\quad\times
  \bigl((a_1+\tfrac{n}{2})^2+(a_2+\tfrac{n}{2})^2
  -(a_3+\tfrac{n}{2}+\ell-1)^2\n
  &\qquad\qquad\qquad\qquad\qquad\qquad
  -(a_4+\tfrac{n}{2}+\ell-1)^2\bigr),\\
  &a_{\ell,n,2}(\bm{\lambda})=
  \prod_{j=1}^2\prod_{k=3}^4(a_j-a_k-\ell+1)\n
  &\quad\times
  \frac{\ell(\ell-1)n(a_3+a_4+2(\ell-1))}
  {(a_1+a_2+n)(a_1+a_2-a_3-a_4-2(\ell-1))},\\
  &b_{\ell,n,1}(\bm{\lambda})=
  -\prod_{j=1}^2\prod_{k=3}^4(a_j+a_k+n+\ell-1)\\
  &\quad\times
  \frac{\ell n(a_1+a_2+n+\ell-1)(a_1+a_2-a_3-a_4-\ell+1)}
  {(a_1+a_2+n)(a_1+a_2+a_3+a_4+2(n+\ell-1))}.
  \nonumber
\end{align}
The new normalisation constant $h_{\ell,n}(\bm{\lambda})$ is
a simple deformation of the original one $h_n(\bm{\lambda})$
\eqref{defhnWilson}:
\begin{align}
  &h_{\ell,n}(\bm{\lambda})=
  \frac{(a_1+a_2+n+\ell)(a_3+a_4+n+2\ell-1)}
  {(a_1+a_2+n)(a_3+a_4+n+\ell-1)}\n
  &\phantom{h_{\ell,n}(\bm{\lambda})=}\times
  h_n(\bm{\lambda}+\ell\bm{\delta}).
\end{align}

\section{Exceptional Askey-Wilson polynomials}
\label{sec:AW}

We present the data in \S\ref{sec:shape} for the Askey-Wilson polynomial.

\subsection{original shape invariant system}
\label{sec:AW_org}

We take the four parameters as follows:
\begin{align}
  &q^{\bm{\lambda}}\eqdef(a_1,a_2,a_3,a_4),\quad
  |a_i|<1\ \ (i=1,\ldots,4),\n
  &\{a_1^*,a_2^*,a_3^*,a_4^*\}=\{a_1,a_2,a_3,a_4\}\ \ (\text{as a set}),
  \label{para_cond_AW}
\end{align}
where $q^{(\lambda_1,\lambda_2,\ldots)}\eqdef
(q^{\lambda_1},q^{\lambda_2},\ldots)$.
We have one more parameter $q$ ($0<q<1$) but its dependence is not
explicitly displayed.
The potential function, ground state, eigenpolynomial, etc. are
given by
\begin{align}
  &V(x;\bm{\lambda})\eqdef
  \frac{\prod_{j=1}^4(1-a_je^{ix})}{(1-e^{2ix})(1-qe^{2ix})},
  \label{VAWilson}\\
  &\phi_0(x;\bm{\lambda})\eqdef
  \biggl|\frac{(e^{2ix};q)_{\infty}}
  {\prod_{j=1}^4(a_je^{ix};q)_{\infty}}\biggr|,\\
  &P_n(\eta(x);\bm{\lambda})=p_n(\cos x;a_1,a_2,a_3,a_4|q)\n
  &\quad
  \eqdef a_1^{-n}(a_1a_2,a_1a_3,a_1a_4;q)_n\\
  &\quad\quad\times
  {}_4\phi_3\Bigl(\genfrac{}{}{0pt}{}{q^{-n},\,a_1a_2a_3a_4q^{n-1},\,
  a_1e^{ix},\,a_1e^{-ix}}{a_1a_2,\,a_1a_3,\,a_1a_4}\Bigm|q;q\Bigr),\n
  &\mathcal{E}_n(\bm{\lambda})=(q^{-n}-1)(1-bq^{n-1}),
  \quad b\eqdef\prod_{j=1}^4a_j,\\
  &f_n(\bm{\lambda})=q^{\frac{n}{2}}(q^{-n}-1)(1-bq^{n-1}),
  \quad b_n(\bm{\lambda})=q^{-\frac{n+1}{2}},
  \label{fbAWilson}\\
  &h_n(\bm{\lambda})=2\pi\,
  \frac{(bq^{n-1};q)_n(bq^{2n};q)_{\infty}}
  {(q^{n+1};q)_{\infty}\prod_{1\leq j<k\leq 4}(a_ja_kq^n;q)_{\infty}}.
  \label{defhnAWilson}
\end{align}
Here $p_n(x;a_1,a_2,a_3,a_4|q)$ is the Askey-Wilson polynomial,
which is symmetric with respect to $a_1,\ldots,a_4$.

\subsection{new shape invariant systems}
\label{sec:AW_new}

We generically restrict the original parameter range
\eqref{para_cond_AW} as follows:
\begin{align}
  &a_1,a_2\in\mathbb{R}, \quad
  \{a_3^*,a_4^*\}=\{a_3,a_4\} \quad (\text{as a set}),\n
  &1>a_j>|a_k|\ \ (j=1,2;k=3,4).
  \label{range:AW}
\end{align}
The polynomial $\xi_{\ell}(x)$ is
\begin{equation}
  \xi_{\ell}(x;\bm{\lambda})=p_{\ell}\bigl(x;
  (a_1q^{\frac{\ell-1}{2}})^{-1},(a_2q^{\frac{\ell-1}{2}})^{-1},
  a_3q^{\frac{\ell-1}{2}},a_4q^{\frac{\ell-1}{2}}|q\bigr),
\end{equation}
which is symmetric under $a_1\leftrightarrow a_2$ and/or
$a_3\leftrightarrow a_4$.
It is again elementary to show that $\xi_{\ell}(x;\bm{\lambda})$ and
$\xi_{\ell}(x;\bm{\lambda}+\bm{\delta})$ have no zero in
the real line segment $0\le x\le \pi$.
The hermiticity of $\mathcal{H}_\ell$ requires a stronger condition
of no zero in the rectangular domain
$-\frac{|\gamma|}{2}\le \text{Im}\, x\le \frac{|\gamma|}{2}$.
It can be determined purely algebraically for each $\ell$.
For the lowest case, $\ell=1$, hermiticity is satisfied if
\begin{align}
  &(a_1+a_2)(1-a_3a_4)-(a_3+a_4)(1-a_1a_2)\n
  &\qquad\qquad\qquad
  >(q^{\frac12}+q^{-\frac12})(a_1a_2-a_3a_4).
  \label{awonerange}
\end{align}
The higher $\ell$ goes the restriction due to hermiticity becomes
the less stringent.

The real coefficients $a_{\ell,n,1}$, $a_{\ell,n,2}$ and $b_{\ell,n,1}$
are
\begin{align}
  &a_{\ell,n,1}(\bm{\lambda})=
  \bigl(q^{n+\ell-1}(a_1+a_2)-q^{n+2(\ell-1)}(a_3+a_4)\n
  &\phantom{a_{\ell,n,1}(\bm{\lambda})=}\quad
  +q^{\ell-1}(a_1^{-1}+a_2^{-1})-(a_3^{-1}+a_4^{-1})\bigr)\\
  &\quad\times
  \frac{q^{\frac{\ell}{2}-1}(1-q^{\ell})(1-q^n)
  (1-a_1^{-1}a_2^{-1}a_3a_4q^{\ell-1})a_3a_4}
  {(1-a_1^{-1}a_2^{-1}a_3a_4q^{2(\ell-1)})
  (1-a_1a_2a_3a_4q^{2(n+\ell-1)})},\n
  &a_{\ell,n,2}(\bm{\lambda})=
  -\prod_{j=1}^2\prod_{k=3}^4(1-a_j^{-1}a_kq^{\ell-1})\n
  &\quad\times
  \frac{(q^{-\ell}-1)(1-q^{\ell-1})(1-q^n)(1-a_3a_4q^{2(\ell-1)})}
  {(1-a_1a_2q^n)(1-a_1^{-1}a_2^{-1}a_3a_4q^{2(\ell-1)})},\\
  &b_{\ell,n,1}(\bm{\lambda})=
  (q^{-\ell}-1)(1-q^n)\prod_{j=1}^2\prod_{k=3}^4(1-a_ja_kq^{n+\ell-1})\n
  &\quad\times
  \frac{(1-a_1a_2q^{n+\ell-1})
  (1-a_1^{-1}a_2^{-1}a_3a_4q^{\ell-1})}
  {(1-a_1a_2q^n)(1-a_1a_2a_3a_4q^{2(n+\ell-1)})}.
\end{align} 
The new normalisation constant $h_{\ell,n}(\bm{\lambda})$ is again
a simple deformation of the original one $h_n(\bm{\lambda})$
\eqref{defhnAWilson}:
\begin{align}
  &h_{\ell,n}(\bm{\lambda})=q^{-\ell}
  \frac{(1-a_1a_2q^{n+\ell})(1-a_3a_4q^{n+2\ell-1})}
  {(1-a_1a_2q^n)(1-a_3a_4q^{n+\ell-1})}\n
  &\phantom{h_{\ell,n}(\bm{\lambda})=}\times
  h_n(\bm{\lambda}+\ell\bm{\delta}).
\end{align}

\subsection{limit to Wilson case}
\label{sec:AWtoW}

It is well known that the Wilson polynomials are obtained from
the Askey-Wilson polynomials in a certain limit \cite{koeswart}
together with the corresponding potential functions, $\mathcal{A}$ and
$\mathcal{A}^\dagger$ operators and the Hamiltonians \cite{os13}.
This is the reason why various formulas for the exceptional
Askey-Wilson polynomials and those of the exceptional Wilson
polynomials are common as shown in \S\ref{sec:shape_new}.
In fact we show that the same limiting procedure will connect
various quantities between the two exceptional polynomials.
Following the  Askey-Wilson $\to$ Wilson limiting arguments in
\cite{os13}, let us introduce the rescaled variable
$x^{\text{W}}=\frac{L}{\pi}x$ and set $\gamma=-\frac{\pi}{L}$,
$q=e^{-\frac{\pi}{L}}$.
This implies
$0<x<\pi\Leftrightarrow 0<x^{\text{W}}<L$,
$p^{\text{W}}=\frac{\pi}{L}p$,
$e^{\gamma p}=e^{-p^{\text{W}}}$.
We write the $x$-dependence explicitly as
$\mathcal{H}_{\ell}=\mathcal{H}_{\ell}(x;\bm{\lambda})$,
$\mathcal{A}_{\ell}=\mathcal{A}_{\ell}(x;\bm{\lambda})$, etc.
By setting
$\bm{\lambda}=(a_1^{\text{W}},a_2^{\text{W}},a_3^{\text{W}},
a_4^{\text{W}})$ ($b=\sum_{j=1}^4a_j^{\text{W}}$),
the desired limit is achieved by $L\to\infty$ ($q\to 1$):
\begin{align}
  &\lim_{L\to\infty}(1-q)^{-3\ell}\xi_{\ell}\bigl(\eta(x);\bm{\lambda}\bigr)
  =\xi_{\ell}^{\text{W}}\bigl(\eta^{\text{W}}(x^{\text{W}});
  \bm{\lambda}\bigr),\\
  &\lim_{L\to\infty}(1-q)^{-3(\ell+n)}
  P_{\ell,n}\bigl(\eta(x);\bm{\lambda}\bigr)
  =P_{\ell,n}^{\text{W}}\bigl(\eta^{\text{W}}(x^{\text{W}});
  \bm{\lambda}\bigr),\\
  &\lim_{L\to\infty}(q;q)_{\infty}^3(1-q)^{3-b+\ell}
  \psi_{\ell}(x;\bm{\lambda})
  =\psi_{\ell}^{\text{W}}(x^{\text{W}};\bm{\lambda}),\\
  &\lim_{L\to\infty}(1-q)^{-2}V_{\ell}(x;\bm{\lambda})
  =V_{\ell}^{\text{W}\,*}(x^{\text{W}};\bm{\lambda}),\\
  &\lim_{L\to\infty}(1-q)^{-1}\varphi(x)
  =\varphi^{\text{W}}(x^{\text{W}}),\\
  &\lim_{L\to\infty}(1-q)^{-1}\mathcal{A}_{\ell}(x;\bm{\lambda})
  =-\mathcal{A}_{\ell}^{\text{W}}(x^{\text{W}};\bm{\lambda}),\\
  &\lim_{L\to\infty}(1-q)^{-2}\mathcal{H}_{\ell}(x;\bm{\lambda})
  =\mathcal{H}_{\ell}^{\text{W}}(x^{\text{W}};\bm{\lambda}),\\
  &\lim_{L\to\infty}(1-q)\mathcal{F}_{\ell}(x;\bm{\lambda})
  =-\mathcal{F}_{\ell}^{\text{W}}(x^{\text{W}};\bm{\lambda}),\\
  &\lim_{L\to\infty}(1-q)^{-3}\mathcal{B}_{\ell}(x;\bm{\lambda})
  =-\mathcal{B}_{\ell}^{\text{W}}(x^{\text{W}};\bm{\lambda}),
  \ \ \text{etc.},
\end{align}
where the superscript ${}^{\text{W}}$ denotes the corresponding
quantity in \S\ref{sec:W_new}.
The orthogonality  \eqref{hln} also correctly reduces to
that of \S\ref{sec:W_new}.

By construction dQM reduces to the ordinary QM in the limit when
the shifts become infinitesimal. Thus the Jacobi polynomials are
obtained from the Askey-Wilson polynomials and the Laguerre polynomials
from the Wilson polynomials in certain limits \cite{koeswart,os6}.
It is possible to write down the limit formulas connecting various
quantities of the $X_\ell$ Askey-Wilson polynomials to those of
the $X_\ell$ Jacobi polynomials, as well as those connecting the
$X_\ell$ Wilson and Laguerre polynomials.
We will defer these rather technical results to a later publication.

\section{Summary and Comments}
\label{sec:summary}

By deforming two well-known shape invariant discrete quantum mechanical
systems, corresponding to the Wilson and Askey-Wilson polynomials,
in terms of their eigenpolynomials, two infinite sets of shape invariant
systems are obtained. Their eigenpolynomials form new types of
orthogonal polynomials, called $X_\ell$ Wilson and Askey-Wilson
polynomials, starting with degree $\ell$ ($\ell=1,2,\ldots$),
which is the degree of the polynomial used for the deformation.
By construction these exceptional orthogonal polynomials do not satisfy
the three term recurrence relation \eqref{threeterm} but they form
complete sets due to the hermiticity (self-adjointness) of the Hamiltonian.
The allowed parameter ranges are determined by the purely algebraic
conditions that the polynomials used for deformation should not have
a zero in the rectangular domain
$x_1\le \text{Re}\, x\le x_2$,
$-\frac{|\gamma|}{2}\le \text{Im}\, x\le \frac{|\gamma|}{2}$.
The explicit ranges for the lowest case
$\ell=1$ are shown in \eqref{wonerange} and \eqref{awonerange}.
The higher $\ell$ goes the restriction due to hermiticity becomes
the less stringent.
It is also possible to find the allowed parameter ranges outside of
those generically specified ranges \eqref{range:W} and \eqref{range:AW}.
It would be interesting to try to perform similar deformation of
other shape invariant systems; for example, the Meixner-Pollaczek,
continuous (dual) Hahn and various reductions of the Askey-Wilson
polynomials.
More interesting would be to formulate the exceptional ($q$-)Racah
polynomials, the most generic examples of discrete quantum mechanics
with real shifts.
It is a good challenge to clarify various properties of these new
polynomials, {\em e.g.\/} generating functions, the Gram-Schmidt
construction, substitutes of the three term recurrence relations
\eqref{threeterm}, etc., and to pursue possible physical applications.

\bigskip
We thank A.\,Khare for useful discussion.
This work is supported in part by Grants-in-Aid for Scientific Research
from the Ministry of Education, Culture, Sports, Science and Technology,
No.19540179.


\end{document}